%% file: sample-sigconf-authordraft.tex
\documentclass[sigconf]{acmart}

\usepackage{xcolor}
\newcommand{\modelname}{GEMs}
\usepackage{makecell}
\usepackage{multirow}
\usepackage{enumitem}

\AtBeginDocument{%
  }

\setcopyright{acmlicensed}
\copyrightyear{2026}
\acmYear{2026}
\acmDOI{XXXXXXX.XXXXXXX}
\acmConference[Conference acronym 'XX]{Make sure to enter the correct
  conference title from your rights confirmation email}{June 03--05,
  2018}{Woodstock, NY}
\acmISBN{978-1-4503-XXXX-X/2018/06}

\begin{document}
\title{\modelname: Breaking the Long-Sequence Barrier in Generative Recommendation with a Multi-Stream Decoder}

\author{Yu Zhou}
\authornotemark[1]
\email{zhouyu12@kuaishou.com}
\affiliation{%
  \institution{Kuaishou Inc.}
  \city{Beijing}
  \country{China}
}
\author{Chengcheng Guo}
\authornotemark[1]
\email{guochengcheng03@kuaishou.com}
\affiliation{%
  \institution{Kuaishou Inc.}
  \city{Beijing}
  \country{China}
}

\author{Kuo Cai}
\email{caikuo@kuaishou.com}
\affiliation{%
  \institution{Kuaishou Inc.}
  \city{Beijing}
  \country{China}
}

\author{Ji Liu}
\email{liuji06@kuaishou.com}
\affiliation{%
  \institution{Kuaishou Inc.}
  \city{Beijing}
  \country{China}
}

\author{Qiang Luo}
\authornotemark[2]
\email{luoqiang@kuaishou.com}
\affiliation{%
  \institution{Kuaishou Inc.}
  \city{Beijing}
  \country{China}
}

\author{Ruiming Tang}
\authornotemark[2]
\email{tangruiming2015@163.com}
\affiliation{%
  \institution{Kuaishou Inc.}
  \city{Beijing}
  \country{China}
}

\author{Han Li}
\email{lihan08@kuaishou.com}
\affiliation{%
  \institution{Kuaishou Inc.}
  \city{Beijing}
  \country{China}
}

\author{Kun Gai}
\email{gai.kun@qq.com}
\affiliation{%
  \institution{Unaffiliated}
  \city{Beijing}
  \country{China}
}

\author{Guorui Zhou}
\email{zhouguorui@kuaishou.com}
\affiliation{%
  \institution{Kuaishou Inc.}
  \city{Beijing}
  \country{China}
}

\thanks{$^\text{*}$ Equal contribution}
\thanks{$^{\dag}$ Corresponding author}

\renewcommand{\shortauthors}{Zhou et al.}

\begin{abstract}
While generative recommendations (GR) possess strong sequential reasoning capabilities, they face significant challenges when processing extremely long user behavior sequences: the high computational cost forces practical sequence lengths to be limited, preventing models from capturing users' lifelong interests; meanwhile, the inherent "recency bias" of attention mechanisms further weakens learning from long-term history. To overcome this bottleneck, we propose \textbf{\modelname} (\underline{\textbf{G}}enerative
r\underline{\textbf{E}}commendation with a \underline{\textbf{M}}ulti-\underline{\textbf{s}}tream decoder), a novel and unified framework designed to break the long-sequence barrier by capturing users' lifelong interaction sequences through a multi-stream perspective. Specifically, {\modelname} partitions user behaviors into three temporal streams—Recent, Mid-term, and Lifecycle—and employs tailored inference schemes for each: a one-stage real-time extractor for immediate dynamics, a  lightweight indexer for cross attention to balance accuracy and cost for mid-term sequences, and a two-stage offline-online compression module for lifelong modeling. These streams are integrated via a parameter-free fusion strategy to enable holistic interest representation. Extensive experiments on  large-scale industrial datasets demonstrate that {\modelname} significantly outperforms state-of-the-art methods in recommendation accuracy. Notably, {\modelname} is the first lifelong GR framework successfully deployed in a high-concurrency industrial environment, achieving superior inference efficiency while processing user sequences of over 100,000 interactions.
\end{abstract}

\ccsdesc[500]{Information systems~Recommender systems}
\keywords{User Interest Modeling,
Lifelong User Behavior,
Generative Recommendation,
Recommendation System
}

\received{14 February 2026}
\received[revised]{14 February 2026}
\received[accepted]{14 February 2026}

\maketitle

\input{chapters/introduction}

\input{chapters/preliminary}
\input{chapters/method2}
\input{chapters/experiment}

\input{chapters/deploy}

\input{chapters/related}

\input{chapters/conclusion}


\bibliographystyle{ACM-Reference-Format}
\bibliography{reference}










\end{document}

%% file: chapters/introduction.tex
\section{Introduction}

In recommendation systems (RS), effective user sequence modeling is crucial for capturing users’ actual intents and enhancing their experience, where the length of the input user sequence is a critical factor.
Empirical studies have consistently shown that scaling the length of user behavior sequences can significantly improve the modeling of long-term user preferences ~\cite{chai2025longerscalinglongsequence}.
Moreover, longer historical sequences can encapsulate richer user information in RS, such as periodic interests and the evolution of preferences over time, thereby enhancing recommendation performance ~\cite{zhai2024actionsspeaklouderwords}. 
In practice, methods such as TWINS ~\cite{Chang2023} and TWINS-V2 ~\cite{Si2024} improve the model's ability to capture user interests by incorporating users' full lifetime sequences, deepening the understanding of user interests and consequently boosting performance in click-through rate (CTR) prediction tasks.
Other works, including MARM ~\cite{Lv2024}, MISS ~\cite{Guo2025}, and ENCODE ~\cite{Zhou2024}, have independently validated the positive impact of modeling users' long-term and even entire lifetime sequences across diverse recommendation scenarios. 
Recently, advances in Large Language Models (LLMs) have opened a promising direction by enhancing user sequence modeling capabilities under scaling laws, leading to the emergence of generative recommendation (GR) as a new paradigm in RS~\cite{202512.0203}. Pioneering methods such as OneRec ~\cite{Zhou2025}, OneRec V2 ~\cite{Zhou2025a}, MGTR ~\cite{Han2025}, PinRec ~\cite{Badrinath2025}, and RecGPT ~\cite{Yi2025} have demonstrated the practical potential of GRs in real-time recommendation tasks. Given the powerful reasoning capabilities of GRs, integrating ultra-long user sequence modeling presents a promising avenue in this emerging field.

\begin{figure}[tbp]
    \centering
    \includegraphics[width=0.95\linewidth]{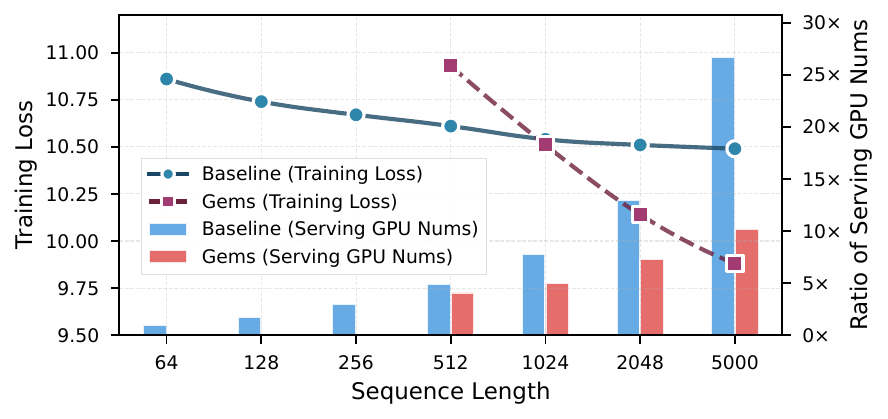} 
    \caption{Analyzing the effect of sequence length on training loss and serving resources. \modelname{} can handle longer user history sequences with fewer inference resources, thereby improving performance on recommendation tasks.}
    \vspace{-0.4cm}
    \label{fig:seqloss}
\end{figure}

Although incorporating lifelong user sequences in GRs has been discussed in several existing works \cite{chen2024hllmenhancingsequentialrecommendations,Han2025,xu2025mmq,yan2025unlockingscalinglawindustrial,zhang2025onetrans}, they have failed to address these critical issues: \begin{itemize}[leftmargin=*]
    \item \textbf{Sequence Length Barrier:} Due to the high computational cost of modeling longer historical sequences in GRs, most industrially deployed GRs are limited to a maximum sequence length of less than 10,000  ~\cite{zhai2024actionsspeaklouderwords,chen2024hllmenhancingsequentialrecommendations,yan2025unlockingscalinglawindustrial,chai2025longerscalinglongsequence}. However, in real-time industrial recommendation scenarios, the most recent 10,000 interaction behaviors typically cover only several months of a user's history, failing to capture their lifelong behavior patterns ~\cite{Si2024}.
    \item \textbf{Inefficient Sequence Scaling:} Empirical experiments indicate that the performance gains achieved by GRs from increasing the length of users' historical sequences are sub-linear. Given the stringent latency requirements for recommendation requests in real-time scenarios, GRs implemented on the Transformer architecture exhibit a low return on investment (ROI) in long-sequence contexts, as illustrated in Fig.~\ref{fig:seqloss}.
    \item \textbf{Recency Bias in Attention:} Concurrently, analyses of the attention mechanisms in GRs from a long-sequence perspective reveal that GRs exhibit a strong tendency to attend primarily to the recent segments of user sequences for inference, which is verified in Sec. ~\ref{exp:fusion}. This pattern hinders their ability to leverage the entirety of long-term user behavioral histories effectively.
\end{itemize}

To tackle these issues, we propose \modelname, a \textbf{unified framework based on GR modeling to capture the entire user lifelong sequences from a multi-stream perspective}. 
Our key insight is to treat long-sequence modeling as a multi-timescale memory reading problem, where different temporal ranges require distinct computational strategies. 
Specifically, {\modelname} partitions user sequences into three streams: \textbf{Recent}, \textbf{Mid-term}, and \textbf{Lifecycle}.
To capture users' short-term interests, which evolve rapidly and reflect recent user dynamics, we employ a one-stage inference scheme for the real-time generation of recent interest representations.
The \textbf{lightweight indexer for cross attention} is specifically applied in mid-term to balance the accuracy of user interest inference with the computational cost of real-time inference.
For the lifecycle stream, we apply a two-stage inference framework: the offline stage processes the users' full lifecycle stream, transforming it into representations that facilitate efficient real-time inference.
The processed representations are then aggregated via a \textbf{parameter-free fusion} strategy—avoiding additional online latency and training instability—before feeding into the unified GR decoder for final recommendation generation.
Based on the aforementioned structure, \modelname{} can handle users’ lifelong sequences with lower inference costs, thereby improving its performance on recommendation tasks.

Our main contributions are:
\begin{itemize}[leftmargin=*]
    \item We propose a unified framework for modeling the entire users' life-time sequence under GR paradigm, allowing GR to efficiently process users' interaction history.
    \item By incorporating users' lifelong historical sequences into GR, we enhance the model's capability to comprehend both long-term and short-term user interests, achieving improved recommendation metrics on industrial-scale datasets and online A/B tests.
    \item We have successfully deployed {\modelname} in industrial settings and optimized its inference efficiency for high-concurrency scenarios through various strategies.
\end{itemize}

%% file: chapters/preliminary.tex
\section{Preliminary}

\textbf{Sequential Recommendations. } Let $\mathcal{X}$ be the entire corpus of items. For user $u$, $X_u=[x_1, ..,x_{t-1}]$ is the interacted sequence of $u$. The sequential recommendation returns a set of item candidates to predict the next interacted item $x_t$.

\textbf{Generative Recommendations. }  
Each item $x$ in the item corpus $\mathcal{X}$ can be represented by an embedding $h \in \mathbb{R}^{dim}$. With a quatinization model such as RQ-VAE and Residual K-means, $h$ can be quantized into a list of discrete codes $[s_1, s_2, \dots, s_d]$, i.e., Semantic IDs.
Generative recommendations take user sequence $X_u=[x_1, ..,x_{t-1}]$ as inputs, then auto-regressively predict the Semantic ID sequence of the next item $x_t$.



%% file: chapters/method2.tex
\begin{figure*}[htbp]
\centering
\includegraphics[width=\textwidth]{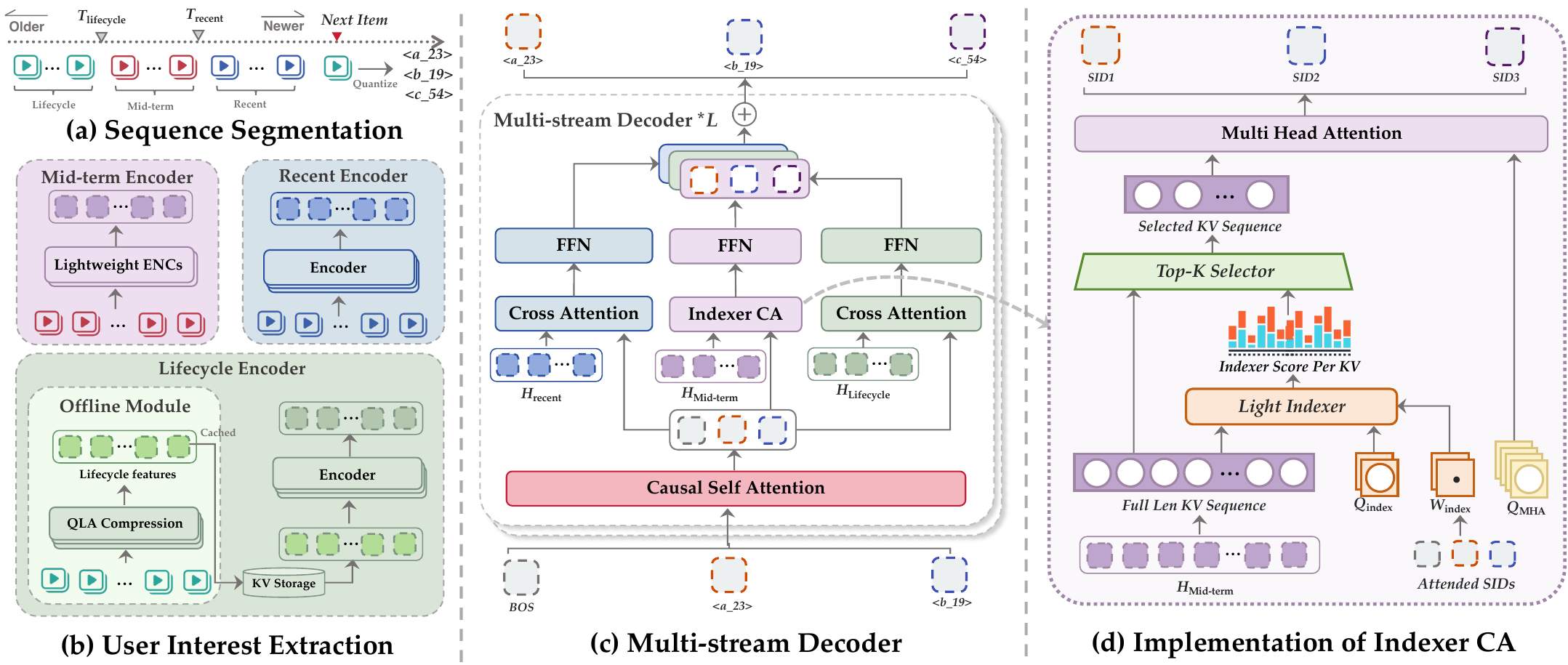}
\caption{Overall framework of \modelname. (a): The user sequence is partitioned iinto three segments—Recent, Mid-term, and Lifecycle. (b): Interest extraction from users' lifelong sequences. We  derive corresponding interest representations through dedicated encoders. (c): A multi-stream decoder for predicting the user’s next item of interest. All streams are then fused via parameter-free fusion to generate the final prediction. (d): Implementation of the Indexer CA block. This module adopts an indexer-selector design that incorporates a pre-filtering strategy on the full-length key-value sequences, significantly reducing the computational cost of Multi-Head Attention.}
\label{figures/overall}
\end{figure*}

\section{Methods}

\subsection{Overview}
\subsubsection{Sequence Segmentation}
The architecture of {\modelname} is shown in Fig.~\ref{figures/overall}. Based on Encoder-Decoder paradigm,  we develop a multi-stream structure to process users' lifelong sequences effectively.

Let the lifelong interaction sequence of user $u$ be defined as $X_u = [x_1, \dots, x_T]$, where $T$ denotes the sequence length. Here, $x_1$ represents the oldest item in the sequence and $x_T$ the most recent one; all subsequent sequence descriptions follow this chronological order from oldest to newest. In real-world industrial deployments, $T$ can be as large as over 100{,}000 interactions.

The sequence is divided based on token budgets, introducing two thresholds $1<T_{\text{lifecycle}}<T_{\text{recent}}<T$. The first threshold, $T_{\text{recent}}$, is used to select the most recent behaviors, which capture users’ immediate feedback. The second threshold, $T_{\text{lifecycle}}$, marks the boundary beyond which behaviors are considered long-term historical interests, covering all possible content a user may have interacted with. The intermediate segment corresponds to mid-term interests.
We maintain a uniform threshold across different users, based on the consideration that user interaction patterns tend to become consistent given a large number of samples.

\begin{equation}\small
\begin{aligned}
    X_{u,\ \text{recent}}&=X_u[T_{\text{recent}}:\ ];\\
    X_{u,\ \text{mid-term}}&=X_u[(T_{\text{lifecycle}}+1):T_{\text{recent}}];\\
    X_{u,\ \text{lifecycle}}&=X_u[:\ T_{\text{lifecycle}}], 
\end{aligned}
\end{equation}
Based on these thresholds, the lifelong interaction sequence ies partitioned into three chronological subsequences: \textbf{Recent}, \textbf{Mid-Term}, and \textbf{Lifecycle}, each representing a distinct stage of user interests, which will be discussed in Sec. \ref{sec:fram}.
We emphasis the implementation of mid-term sequences using a lightweight indexer mechanism detailed in Section~\ref{method/mid}. Subsequently, a parameter-free fusion strategy to integrate the user’s sub-interest representations is introduced in Section~\ref{method/multi}. 
Finally, the training and inference strategy is presented in Section~\ref{method/train}.

\subsubsection{Multi-Stream Framework}\label{sec:fram}
To effectively process sequences of vastly different lengths and characteristics, we design specialized encoders for each stream:

\textbf{Recent Stream.} For the recent sequence $X_{u,\ \text{recent}}$, we extract multi-modal features including video ID ($vid$), author ID ($aid$), video tags ($tag$), timestamp ($ts$), playtime ($pt$), video duration ($dur$), and interaction labels ($label$). These features are concatenated and projected via an MLP to obtain hidden representations:

\begin{equation}\small
\begin{aligned}
    f_{\text{recent}}^{i} &= [e_{vid}^{i}, e_{aid}^{i}, e_{tag}^{i}, e_{ts}^{i}, e_{pt}^{i}, e_{dur}^{i}, e_{label}^{i}]\\
    h_{recent}^{i} &= \operatorname{MLP}(f_{\text{recent}}^{i}),\ i \in [1, T_{\text{recent}}]
\end{aligned}
\end{equation}
We then apply multi-layer self-attention with position embeddings to capture users' immediate interests from hidden representations.

\textbf{Mid-term Stream.} For mid-term sequences containing thousands of tokens, we employ a lightweight indexer-based cross attention mechanism. This approach selectively attends to informative tokens through a sparse attention scheme, significantly reducing computational complexity while maintaining signal quality. Section~\ref{method/mid} details the implementation of the lightweight indexer.

\textbf{Lifecycle Stream.} To handle ultra-long sequences exceeding 100,000 interactions, inspired by VISTA~\cite{chen2025massive}, we adopt a two-stage mechanism. In the offline stage, we compress the lifecycle history using a quasi-linear attention (QLA) with QLU optimization:

\begin{equation}\small
\begin{aligned}
    \operatorname{QLUAttn}(T, E) =\varphi(Q[E]) \varphi\left(\varphi(K[T])^{\top} V[T]\right)+ \\
    \Delta(\varphi(Q[E]), \varphi(K[T])) V[T]
\end{aligned}
\end{equation}
This compression is supervised via a generative reconstruction loss $\mathcal{L}_{\text{recon}} = \sum_{i=1}^{M-1}\left\|v_i-u_{i+1}\right\|_2^2$, where $v_i$ and $u_{i+1}$ represent compressed and reconstructed representations. During inference, compressed representations are retrieved from a low-latency KV system and processed by a lightweight self-attention module.

\subsection{Lightweight Indexer for Cross Attention} \label{method/mid}
We propose a \textbf{lightweight indexer} mechanism for modeling our \textbf{mid-term sequences}.
Accurately capturing interest shifts beyond recent preferences can improve recommendation diversity and reveal interests underrepresented in recent behavior.
However, mid-term sequences typically contain thousands of tokens, making full attention computationally expensive and noisy.
To address this, we propose a lightweight indexer for cross attention that selectively attends to informative tokens, thereby enhancing the signal-to-noise ratio and reducing complexity.

\textbf{Lightweight Indexer.} 
To mitigate the computational cost of full cross-attention, we introduce a lightweight indexer that selects a high-value subset of tokens. The indexer is trained to mimic the attention distribution of full cross-attention blocks but uses far fewer heads.
The mid-term sequence $X_{u, \text{mid-term}}$ is processed into initial hidden representations $H^{(0)}{\text{mid-term}}$ using the aforementioned behavior encoder followed by an MLP projection. The subsequent decoder comprises $L{\text{mid}}$ layers, each with $H$ standard attention heads.
For the $h$-th head in the $j$-th block, during training, we first compute the full cross-attention scores $S^{(j,h)}{\text{mid-term}}$ :
\begin{equation}\small
\begin{aligned}
    Q^{(j,h)}_{\text{mid-term}}&=W^{(j,h)}_{Q, \text{mid-term}}H^{(j-1)}_{\text{mid-term}},\\
    K^{(j)}_{\text{mid-term}}&=W^{(j,h)}_{K, \text{mid-term}}H^{(j-1)}_{\text{mid-term}},\\
    S^{(j,h)}_{\text{mid-term}}&={Q^{(j,h)}_{\text{mid-term}}}^TK^{(j,h)}_{\text{mid-term}}/\sqrt{d_k},\\
    S^{(j)}_{\text{mid-term}}&=\sum_{h=1}^HS^{(j,h)}_{\text{mid-term}}.
\end{aligned}
\end{equation}
We then compute an alternative attention score using a separate projection with far fewer heads $H_{\text{indexer}} \ll H$:
\begin{equation}\small
\begin{aligned}
    Q^{(j,h)}_{\text{indexer}}&=W^{(j,h)}_{Q, \text{indexer}}H^{(j-1)}_{\text{mid-term}},\\
    K^{(j)}_{\text{indexer}}&=W^{(j,h)}_{K, \text{indexer}}H^{(j-1)}_{\text{mid-term}},\\
    S^{(j,h)}_{\text{indexer}}&={Q^{(j,h)}_{\text{indexer}}}^TK^{(j,h)}_{\text{indexer}}/\sqrt{d_k},\\
    D^{(j)}&=W^{(j)}_{D}H^{(j-1)}_{\text{mid-term}},\\
    S^{(j)}_{\text{indexer}}&={D^{(j)}}^T[S^{(j,1)}_{\text{indexer}};...;S^{(j,H_{\text{indexer}})}_{\text{indexer}}].
\end{aligned}
\end{equation}
Here $S^{(j)}{\text{indexer}}$ denotes the scoring output of the lightweight indexer at block $j$.
It is aligned with the full attention score $S^{(j)}{\text{mid-term}}$ via Kullback–Leibler divergence:
\begin{equation}\small
    \mathcal{L}_{indexer}=\sum_{j=1}^LD_{KL}(S^{(j)}_{\text{indexer}}\ |\ \text{Softmax}(S^{(j)}_{\text{mid-term}}))
\end{equation}

\textbf{Sparse Attention Mechanism.}
The indexer scores reflect the importance of each key with respect to a given query.
Based on the trained $S^{(j)}{\text{indexer}}$, we select the top-$K$ keys and values per query to perform sparse attention.
Each block also includes residual connections and a feed-forward network:
\begin{equation}\small
\begin{aligned}
    H^{(j)\prime}_{\text{mid-term}}&=H^{(j-1)}_{\text{mid-term}}+\text{SparseCrossAttn}(H^{(j-1)}_{\text{mid-term}},\ i \in \text{Top}K(S^{(j)}_{\text{indexer}})),\\
    H^{(j)}_{\text{mid-term}}&=H^{(j)\prime}_{\text{mid-term}}+\text{FFN}(\text{RMSN}(H^{(j)}{'}_{\text{mid-term}})).
\end{aligned}
\end{equation}
$H^{(L_{\text{mid}})}_{\text{mid-term}}$ is the output of the indexer-based cross-attention.
During inference, selecting high-quality tokens via the indexer eliminates full attention over all tokens, shifting part of the computational cost to training.
Our experiments include comprehensive comparisons to validate the inference efficiency of this sparse attention scheme.

\subsection{Parameter-Free Fusion}\label{method/multi}

The decoder  autoregressively generate the Semantic IDs for the user's next item. We take the representation vector of the target video $x_t$'s Semantic IDs, $[E_{s^{\text{1}}_{x_t}};...;\ E_{s^{\text{d}}_{x_t}}]$:
\begin{equation}\small
    D^{(0)}=[E_{\text{BOS}};\ E_{s^{\text{1}}_{x_t}};...;\ E_{s^{\text{d}}_{x_t}}]
\end{equation}
Assuming there are $L$ blocks, each decoder block $j$ first applies multi-head causal self-attention, allowing tokens at a later hierarchical level to gather information from preceding SID tokens:
\begin{equation}\small
    D^{(j)\prime} = D^{(j-1)}+\text{CausalSelfAttn}(D^{(j-1)}),
\end{equation}
Subsequently, $D^{(j)\prime}$ interacts with user interests via stream-specific cross-attention over the encoded memories.
Each stream has independent parameters, and both are equipped with residual connections.
For stream $s \in \{\text{recent},\ \text{mid-term},\ \text{lifecycle}\}$, we compute:

\begin{equation}\small
\begin{aligned}
    D^{(j)}_{s} &= D^{(j)\prime} + \mathrm{CrossAttn}\!\left(D^{(j)\prime},\ H^{(L_s)}_{s},\ H^{(L_s)}_{s}\right),\\
    D^{(j)*}_{s} &= D^{(j)}_{s} + \mathrm{FFN}\!\left(\mathrm{RMSN}\!\left(D^{(j)}_{s}\right)\right).
\end{aligned}
\end{equation}

Three hidden states for each stream are formed: $D^{(j)}_{\text{recent}}$, $D^{(j)}_{\text{mid-term}}$, and $D^{(j)*}_{\text{lifelong}}$.  The method for fusing multiple user sequence features on the decoder is crucial for final generation quality. Interestingly, the most effective approach is also the simplest. We directly fuse hidden states using a non-learnable, parameter-free summation:
\begin{equation}\small
    D^{(j)}=D^{(j)*}_{\text{recent}}+D^{(j)*}_{\text{mid-term}}+D^{(j)*}_{\text{lifelong}}.
\end{equation}




In practice, many prior works fuse multiple sequences using learnable weights, like gating or MLP-based fusion strategy, which can be effective when different streams are drawn from similar distributions. 
In our ultra-long sequence setting, however, the mid-term and lifecycle streams exhibit a larger distribution shift from the next-item target than the recent stream. 
As a result, learnable fusion often collapses to over-weighting the recent stream, weakening the contribution of longer-term signals. 
We validate this effect empirically in Section~\ref{exp:fusion} and provide a visualization analysis.

\begin{table*}[!ht]
\centering
\caption{Overall performance of {\modelname} and baselines evaluated by $\text{Recall}@K$ and $\text{NDCG}@K$, where $K\in\{100,500,1000\}$. The best results are in \textbf{bold}, and the second-best are \underline{underlined}. {\modelname} shows promising results by outperforming all baselines.}
\begin{tabular}{lccccccc}
\toprule
\multicolumn{1}{c}{\textbf{Methods}} & \multicolumn{1}{c}{\textbf{Seq Len}} & \multicolumn{1}{c}{\textbf{Recall@100}} & \multicolumn{1}{c}{\textbf{NDCG@100}} & \multicolumn{1}{c}{\textbf{Recall@500}} & \multicolumn{1}{c}{\textbf{NDCG@500}} & \multicolumn{1}{c}{\textbf{Recall@1000}} & \multicolumn{1}{c}{\textbf{NDCG@1000}} \\
\midrule
MPFormer & 64  & 0.0706 & 0.0033 & 0.1581 & 0.0069 & 0.2010 & 0.0084 \\
         & 256 & 0.0803 & 0.0037 & 0.1794 & 0.0073 & 0.2114 & 0.0090 \\
TDM      & 2000 & 0.0674 & 0.0027 & 0.1677 & 0.0071 & 0.2098 & 0.0084 \\
         & 4000 & 0.0635 & 0.0026 & 0.1701 & 0.0072 & 0.2112 & 0.0086 \\
MISS     & 4000 & 0.0844 & 0.0028 & 0.2033 & 0.0074 & 0.2441 & 0.0089 \\
GPRP     & N/A    & 0.0414 & 0.0015 & 0.0661 & 0.0030 & 0.0929 & 0.0042 \\
Kuaiformer & 64  & 0.0622 & 0.0029 & 0.1388 & 0.0063 & 0.1761 & 0.0080 \\
         & 256 & 0.0724 & 0.0031 & 0.1421 & 0.0071 & 0.1901 & 0.0081 \\
MIND     & 1000 & 0.0457 & 0.0024 & 0.0933 & 0.0049 & 0.1445 & 0.0069 \\
         & 2000 & 0.0481  & 0.0025   & 0.1012  & 0.0051  & 0.1501 & 0.0070 \\
CRM      & 64   & 0.0409 & 0.0022 & 0.0977 & 0.0049 & 0.1437 & 0.0068 \\
         & 256  & 0.0591 & 0.0028 & 0.1034 & 0.0051 & 0.1513 & 0.0071 \\
TIGER    & 256  & 0.1017 & 0.0044 & 0.1914 & 0.0101 & 0.2350 & 0.0120 \\
         & 1024 & \underline{0.1282} & \underline{0.0053} & 0.2351 & 0.0104 & 0.2847 & 0.0139 \\
GRank    & 1024 & 0.1010  & 0.0047 & \underline{0.2396} & \underline{0.0109} & \underline{0.3178} & \underline{0.0142} \\
\midrule
\textbf{Proposed} & \textbf{Lifelong} & \textbf{0.1553} & \textbf{0.0064} & \textbf{0.2911} & \textbf{0.0122} & \textbf{0.3511} & \textbf{0.0145} \\
\midrule
\textbf{Improv.} & \textbf{} & \textbf{21.14\%} & \textbf{20.75\%} & \textbf{21.49\%} & \textbf{11.93\%} & \textbf{10.47\%} & \textbf{2.11\%} \\
\bottomrule
\end{tabular}
\label{tab:results}
\end{table*}

\subsection{Training and Inference Strategy}\label{method/train}

\textbf{Training Stage}
The main task is optimized using $\mathcal{L}_{\mathrm{NTP}}$: the final hidden state $D^{(j)}$ of the decoder is dot-producted with codebook vectors $\mathbf{C} \in \mathbb{R}^{M \times d_h}$ to compute the generation probability $p\theta(s_{x_t}^d \mid \ldots)$ of the positive sample. Negative log-likelihood is then used to maximize this probability:
\begin{equation}\small
    \mathcal{L}_{\mathrm{NTP}} = -\log p_\theta(s_{x_t}^d \mid \ldots).
\end{equation}
The overall training loss for the model is defined as:
\begin{equation}\small
    \mathcal{L}=\mathcal{L}_{\mathrm{NTP}} + \lambda\mathcal{L}_{indexer},
\end{equation}where $\lambda \in \{1, 0\}$ controls whether $\mathcal{L}_{\text{indexer}}$ are back propagated.

The training of {\modelname} consists of three stages: 
\begin{itemize}[leftmargin=*]
\item \textbf{Stage 1: Full-sequence training convergence.} Token selection by the indexer is disabled, and gradients from $\mathcal{L}_{\text{indexer}}$ are frozen ($\lambda = 0$). This stage primarily converges the original item tokens and full-attention parameters, preventing potential instability in the indexer's supervision signal.
\item \textbf{Stage 2: Indexer training convergence.} After $\mathcal{L}_{\mathrm{NTP}}$ converges, $\mathcal{L}_{\text{indexer}}$ are enabled ($\lambda = 1$) to align $S^{(j)}_{\text{indexer}}$ with $S^{(j)}_{\text{mid-term}}$.
 \item \textbf{Stage 3: Indexer inference convergence.} Once $\mathcal{L}_{\text{indexer}}$ converges, token selection via the indexer is activated, ensuring that the final training objective aligns with the online inference task.
\end{itemize}
It is noteworthy that the indexer is detached during training. The indexer is optimized solely based on $\mathcal{L}_{\text{indexer}}$. This design decouples the 'select' and 'attention' operations in the sparse attention mechanism, preventing the leakage of the indexer's reasoning capabilities into the cross-attention module. This safeguards against potential discrepancies between the indexer's training and inference stages.

\textbf{Inference Stage}
We incorporate a beam search structure during the inference stage to facilitate end-to-end recommendation.
Specifically, 
during the inference stage of {\modelname}, beginning with the $[BOS]$ token, {\modelname} generates next token through a beam-search strategy.
The number of generation steps equals the codebook depth, while the beam width at each level is determined by the number of items to be returned per request.

%% file: chapters/experiment.tex
\begin{table}[!ht]
    \caption{Dataset statistics.}\vspace{-0.1cm}
    \centering
    \label{tab:stat}
      \resizebox{.75\linewidth}{!}{\begin{tabular}{ll}
            \toprule
        \textbf{Data} & \textbf{Size}\\
             \midrule
        Users  & 400 million per day \\
        Items & 100 million per day \\
        User-item Interactions & 50 billion per day \\
        Avg. User Sequence Length & 14.5 thousand \\
        Max. User Sequence Length & 100 thousand \\
        \bottomrule
    \end{tabular}}
\end{table}

\section{Experiments}
In this section, extensive empirical experiments are conducted with {\modelname} to answer the following research questions:
\\\textbf{RQ1:} How does {\modelname} perform compared to existing strong baselines for sequential modeling?
\\\textbf{RQ2:} Does each introduced stream increase the final performance?
\\\textbf{RQ3:} Is the proposed parameter-free fusion mechanism optimal compared to other fusion strategies?
\\\textbf{RQ4:} Is the proposed indexer-based encoder structure efficient?
\\\textbf{RQ5:} How does the model size influence the experimental results?
\\\textbf{RQ6:} Does the model lead to improvements in real-world industry A/B testing metrics?

\subsection{Dataset} In related work on ultra-long user sequence modeling, real industrial datasets rather than public benchmarks are commonly used for evaluation \cite{Chang2023, Si2024, chai2025longerscalinglongsequence}. The primary reason is that existing public benchmarks are limited in scale and lack sufficient user historical behaviors. For example, the average sequence length in the Amazon Review \cite{mcauley2015image} dataset is about 8, while sequences in the Taobao \cite{pi2019practice} dataset are mostly around 500. Even the longest dataset \cite{harper2015movielens} we identified in our survey contains sequences only at the thousand-level. Achieving good experimental results on these datasets does not necessarily translate to industrial scenarios, especially where user lifetime sequences often exceed 100,000 interactions. Therefore, following the practice established in prior work on ultra-long user sequence modeling, we conduct experiments on real-world data from an industrial application.

Our experiments are performed on data from Kuaishou, one of the leading short-video  platforms in China. Relevant dataset statistics are provided in Tab. \ref{tab:stat}. The training logs cover 400 million daily active users, 100 million items and 50 billion logs each day. We train models (including baselines) in online learning manners. 
        
\subsection{Metrics} For overall performance, we use $\text{Recall}@K$ and $\text{NDCG}@K$ ($K\in\{100,500,1000\}$) to evaluate model performance.

In addition, we also incorporate the hierarchical recall ($\text{Hrecall}@$ $\text{Level}l@1000$) proposed by \citet{guo2026promiseprocessrewardmodels} as an additional evaluation measure. This metric assesses the recall at any intermediate step during the beam search process, which is crucial for evaluating model performance at different stages of inference. Consequently, we focus on this metric in our ablation studies to verify the contribution of each proposed stream to the final results.

\subsection{Overall Performance (RQ1)}

\subsubsection{Baselines.} We include 9 methods for sequential modeling as our baselines:
\begin{itemize}[leftmargin=*]
    \item \textbf{MPFormer} \cite{sun2025mpformer}: A transformer-based multi-target retrieval.
    \item \textbf{TDM} \cite{zhu2018learning}: A tree-based deep model for large-scale retrieval. 
    \item \textbf{MISS} \cite{Guo2025}: A multi-modal tree-based retrieval with bi-stream sequence modeling.
    \item \textbf{GPRP} \cite{zheng2024full}: A retrieval model using cascaded  samples.
    \item \textbf{Kuaiformer} \cite{liu2024kuaiformer}: A transformer-based retrieval model. 
    \item \textbf{MIND} \cite{li2019multi}: A capsule network for multi-interest modeling.
    \item \textbf{TIGER} \cite{rajput2023recommender}: An encoder-decoder structure for SID generation.
    \item \textbf{GRank} \cite{sun2025grank}: A retrieve-then-rank model for recommendation.
    \item \textbf{CRM} \cite{liu2024crm}: A transformer-based controllable retrieval model.
\end{itemize}

\subsubsection{Results.}
\begin{itemize}[leftmargin=*]
\item \textbf{{\modelname} outperforms all baseline methods.} Across all evaluated cut-offs $K\in{100,500,1000}$, {\modelname} achieves superior performance over all baselines in terms of both $\text{Recall}@K$ and $\text{NDCG}@K$. Specifically, it yields a relative improvement of 21.14\% in $\text{Recall}@100$ and 21.21\% in $\text{NDCG}@100$ compared to the best baseline. Similarly, improvements of 21.49\% in $\text{Recall}@500$ and 8.28\% in $\text{NDCG}@500$ are observed.

\item \textbf{{\modelname} models significantly longer sequences.} Among the baselines, MISS models sequences up to 4000 in length using two SIM modules, while GRank handles sequences of 1024 via attention mechanisms. By contrast, \modelname\ incorporates lifelong-length sequences,  contributing to its recommendation performance.

\item \textbf{The relative improvement of {\modelname} is more pronounced at smaller retrieval sizes.} For instance, at $K=100$, the relative gains in $\text{Recall}@100$ and $\text{NDCG}@100$ are notably larger compared to those at higher values of $K$.

\end{itemize}

\begin{figure*}[tb]
    \centering
    \includegraphics[width=\textwidth]{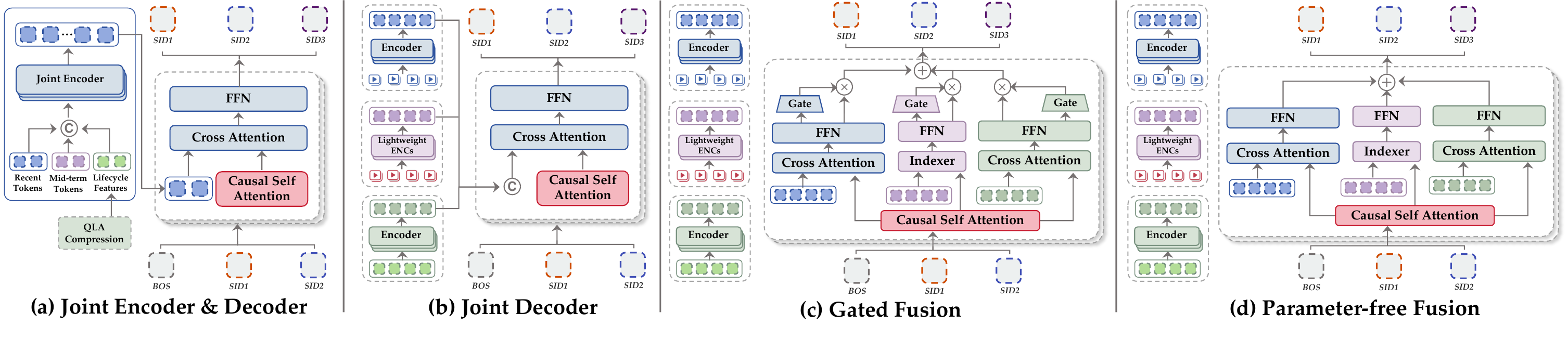}\vspace{-0.4cm}
    \caption{Comparison of fusion strategies. (a): All streams share a single encoder and a single decoder.  (b): Separate encoders are used for the three streams, and their concatenated outputs are fed into a joint decoder. (c): Separate encoders and cross-attentions, whose outputs are fused fused with gated weighting. (d): Our proposed parameter-free fusion strategy, where each stream employs independent cross-attentions, whose outputs are fused without learnable parameters.  }
    \label{fig:fusion} 
\end{figure*}

\subsubsection{Implementation Details.}
In {\modelname}, the hidden dimension is set to 1024. The maximum user sequence length can be as long as $T=100,000$. We define the most recent $T_{\text{recent}}=256$ interactions as the short-term sequence. Interactions from $T_{\text{recent}}$ up to $T_{\text{lifecycle}}=5,000$ are designated as the mid-term sequence, and all interactions beyond that constitute the lifelong sequence. All items are quantized using Residual K-means into a three-level codebook with sizes $[8192, 8192, 8192]$. The decoder consists of 4 layers, and the recent-stream also employs 4 layers. The standard attention mechanism uses 8 heads, while the Lightweight Indexer uses 2 attention heads.

\subsection{Ablation Studies (RQ2)}

\begin{table}[!htb]
    \centering
    \caption{The ablation results of removing streams evaluated by Hrecall@Level\textit{l}@1000. The best results are in \textbf{bold}. Removing any stream leads to performance degradation.
    }\label{tab:abl}
    \resizebox{\linewidth}{!}{\begin{tabular}{llccc}
            \toprule
        \textbf{ID} & \textbf{Settings} & \textbf{\makecell{Hrecall@\\Level1@1000}} & \textbf{\makecell{Hrecall@\\Level2@1000}} & \textbf{\makecell{Hrecall@\\Level3@1000}}  \\ \midrule
        {\small(1)} &Recent only & 0.9311 & 0.4093 & 0.2660  \\ 
        {\small(2)} &Recent and mid-term only & 0.9372 & 0.4322 & 0.3310  \\
        {\small(3)} &Recent and lifecycle only & 0.9420 & 0.4290 & 0.3210  \\ 
        {\small(4)} &\textbf{Proposed} & \textbf{0.9420} & \textbf{0.5115} & \textbf{0.3817} \\ \bottomrule
    \end{tabular}}
\end{table}
This work integrates recent, mid-term, and lifecycle streams. In this experiment, we aim to verify that each stream introduced by the proposed model contributes positively to performance by ablating specific streams from the full model. The experimental results are presented in Table \ref{tab:abl}.

Row (4) corresponds to the proposed full architecture. In row (3), we remove the lifecycle stream. While the hierarchical recall at Level 1 remains comparable, it drops substantially at subsequent levels, indicating that the lifecycle structure contributes significantly to the later stages of inference. In row (2), we remove the mid-term stream, which leads to a notable drop in performance at Level 1. The degradation at deeper levels, though present, is less severe than in row (3), suggesting that the mid-term stream is particularly beneficial for first-level inference. Finally, in row (1), where only the recent stream is retained, performance declines significantly across all levels. These results collectively demonstrate the effectiveness and rationale behind our multi-stream design.

\subsection{Analysis of Fusion Strategies (RQ3)}\label{exp:fusion}
\begin{table}[htbp]
    \centering
    \caption{Fusion strategies evaluated by Hrecall@Level\textit{l}@1000. The best results are in \textbf{bold}. An illustration of  strategy (a)—(d) is shown in Fig. \ref{fig:fusion}.  }\label{tab:fusion}
    \resizebox{.85\linewidth}{!}{\begin{tabular}{cccc}\toprule
  \textbf{\makecell{Fusion\\Statrategy}} & \textbf{\makecell{Hrecall@\\Level1@1000}} & \textbf{\makecell{Hrecall@\\Level2@1000}} & \textbf{\makecell{Hrecall@\\Level3@1000}} \\
    \midrule
    {\small(a)} & 0.9317 & 0.4750 & 0.3412 \\
    {\small(b)} & 0.9368 & 0.4834 & 0.3603 \\
    {\small(c)} & 0.9360 & 0.4819 & 0.3537\\ 
    {\small(d)} & \textbf{0.9420} & \textbf{0.5115} & \textbf{0.3817} \\\bottomrule
  \end{tabular}}
\end{table}

\begin{figure}[htbp]
    \centering\vspace{-0.3cm}
    \includegraphics[width=\linewidth]{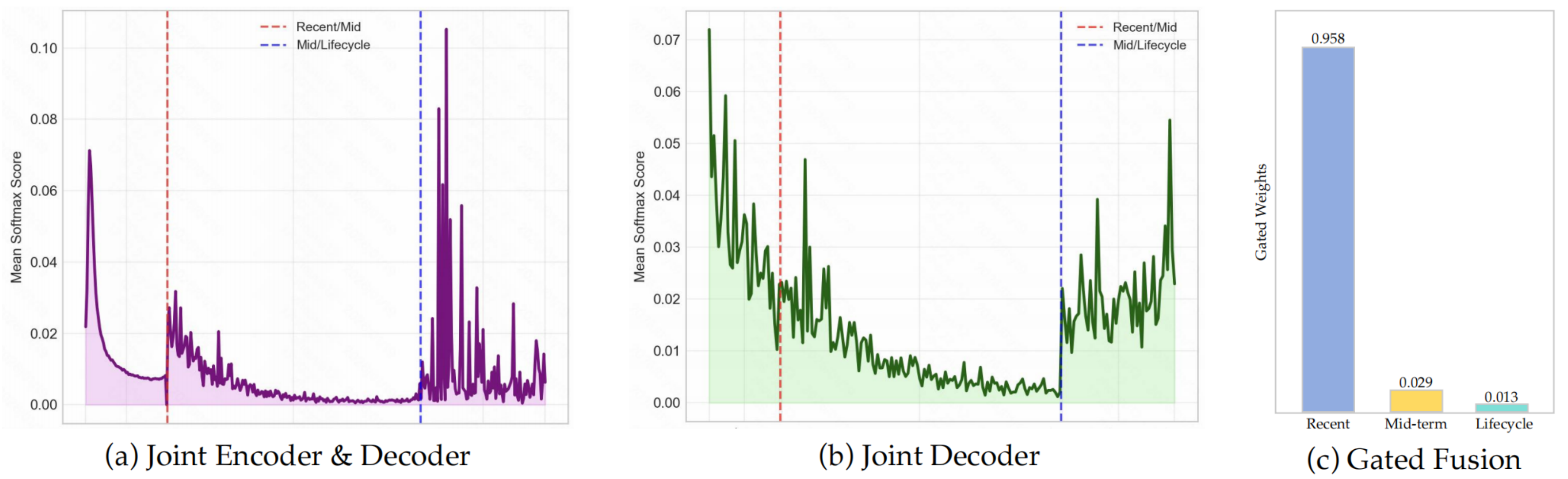}
    \caption{Analysis of fusion stageties. The model tends to
     allocate more attention to the recent stream.}\vspace{-0.3cm}
    \label{fig:gate} \vspace{-0.2cm}
\end{figure}

We propose a parameter-free fusion strategy for our multi-stream decoder, which we argue is optimal for modeling ultra-long sequences in real-world industrial scenarios. Therefore, we compare the optimal strategy (d), i.e., the parameter-free fusion strategy, with three alternative strategies. 
\begin{itemize}[leftmargin=*]
    \item {(a)} Full concatenation of input tokens from all streams, processed by a single shared encoder and decoder;
    \item {(b)} Separate encoders for each stream, with their outputs concatenated and fed into a shared decoder;
    \item {(c)} A variant identical to (d) except that the three streams are fused using gated weighting.
\end{itemize}
The corresponding architectures are illustrated in Fig. \ref{fig:fusion}. Empirical results are presented in Table \ref{tab:fusion}.

Our parameter-free fusion strategy achieves the best experimental results. This can be attributed to the fact that the parameter-free fusion strategy does not employ learnable parameters to weight the different streams. When learnable parameters are used to assign weights to the streams, the model tends to favor the most recent stream, failing to learn from earlier streams. This behavior is empirically supported by Fig. \ref{fig:gate}, where we visualize the cross-attention weights across the three streams in strategies (a) and (b). As shown, \textbf{the majority of attention is allocated solely to the recent stream}.
In contrast, the parameter-free fusion strategy, by avoiding learnable weighting parameters, allows all three streams to contribute equally to the final recommendation outcome.
We also visualize strategy (c) with gated weighting in Fig. \ref{fig:gate}, demonstrating that adding learnable parameters impairs the model’s ability to capture information equally from sequences across different timescales, which consequently leads to poorer results in Table \ref{tab:abl}.

\subsection{Analysis of Indexer Efficiency (RQ4)}
\begin{table}[!ht]
    \centering\caption{Comparison between using the indexer and brute-force extraction for the mid-term sequence, in terms of inference latency and performance. The best results are highlighted in \textbf{bold}.\label{tab:ie}
    }
    \resizebox{0.8\linewidth}{!}{\begin{tabular}{lllcc}
    \toprule
        \multicolumn{3}{c}{\textbf{Settings}} &  \multicolumn{2}{c}{\textbf{Performance}}    \\ 
        \cmidrule(lr){1-3} \cmidrule(lr){4-5}
        \textbf{\makecell[c]{Aggregation\\Method}} & \textbf{\textit{T}}$\mathbf{_{recent}}$ & \textbf{\textit{T}}$\mathbf{_{lifecycle}}$ & \textbf{\makecell{Inference\\Latency (ms)}} & \textbf{\makecell{Recall\\@1000}}  \\ \midrule
        Brute-force & 1000 & 2000 & 60.18 & 0.3312  \\ 
        Brute-force & 1000 & 5000 & 103.68 & 0.3464  \\ 
        Indexer & 1000 & 5000 & \textbf{32.2} & \textbf{0.3511} \\ \bottomrule
    \end{tabular}}
\end{table}

We propose a novel lightweight indexer for cross attention to efficiently handle the mid-term sequence while reducing computational complexity. In this section, we empirically investigate the efficiency of the indexer. Specifically, we compare our lightweight indexer with a brute-force self-attention mechanism applied to the mid-term sequence. This comparison evaluates whether the indexer is computationally efficient and whether it can mitigate noise in the mid-term sequence to improve recommendation performance.

The results are shown in Table \ref{tab:ie}. The brute-force approach significantly increases inference latency. With \(T_{\text{lifecycle}} = 5000\), the latency increases by approximately fivefold. Even when the sequence length is reduced, the latency remains higher than that of the indexer. Furthermore, the indexer's ability to filter sequence noise leads to better performance than the brute-force method. These findings demonstrate that our designed lightweight indexer improves performance while achieving lower inference latency.

\subsection{Analysis of Model Size (RQ5)}

\begin{figure}[htbp]\vspace{-0.3cm}
    \centering
    \includegraphics[width=0.8\linewidth]{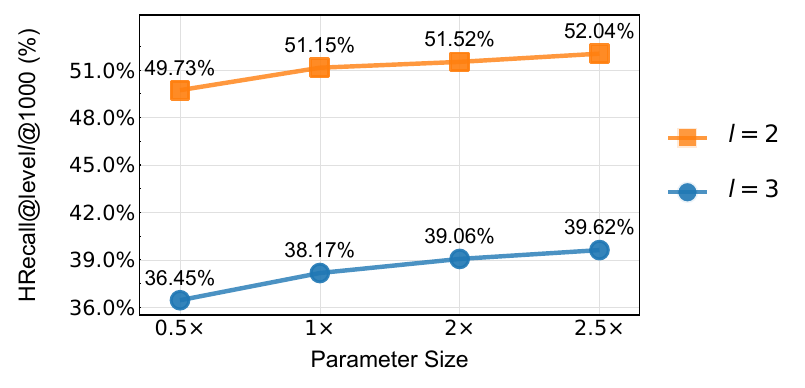}\vspace{-0.3cm}
    \caption{Analysis of model size. The model's capacity increases with scaling.}
    \label{fig:md_sz} \vspace{-0.2cm}
\end{figure}

\begin{table*}[!ht]
    \centering\caption{Online A/B test results. This table shows relative improvements of experimental groups over control groups. All results are statistically significant on 0.05 significant level. \label{tab:ab}
    }
\resizebox{0.8\linewidth}{!}{\begin{tabular}{lcccccc}
\toprule
\textbf{Apps}          & \multicolumn{1}{c}{\textbf{\begin{tabular}[c]{@{}c@{}}Total App\\      Usage Time\end{tabular}}} & \multicolumn{1}{c}{\textbf{\begin{tabular}[c]{@{}c@{}}Total App\\      Usage Time (CI)\end{tabular}}} & \multicolumn{1}{c}{\textbf{\begin{tabular}[c]{@{}c@{}}App Usage Time\\      per User\end{tabular}}} & \multicolumn{1}{c}{\textbf{\begin{tabular}[c]{@{}c@{}}App Usage Time\\      Per User (CI)\end{tabular}}} & \multicolumn{1}{c}{\textbf{\begin{tabular}[c]{@{}c@{}}Video\\      Watch Time\end{tabular}}} & \multicolumn{1}{c}{\textbf{\begin{tabular}[c]{@{}c@{}}Watch Time\\      per Video View\end{tabular}}} \\ \midrule
Kuaishou      & \multicolumn{1}{c}{0.13\%}                                                              & \multicolumn{1}{c}{{[}+0.07\%, +0.19\%{]}}                                                   & \multicolumn{1}{c}{0.11\%}                                                                 & \multicolumn{1}{c}{{[}+0.05\%, +0.17\%{]}}                                                      & \multicolumn{1}{c}{0.20\%}                                                          & \multicolumn{1}{c}{0.42\%}                                                                   \\
Kuaishou Lite & \multicolumn{1}{c}{0.17\%}                                                              & \multicolumn{1}{c}{{[}+0.06\%,+ 0.27\%{]}}                                                   & \multicolumn{1}{c}{0.14\%}                                                                 & \multicolumn{1}{c}{{[}+0.07\%,+ 0.25\%{]}}                                                      & \multicolumn{1}{c}{0.35\%}                                                          & \multicolumn{1}{c}{0.26\%}                                                                   \\
                                                                            \bottomrule             
\end{tabular}}
\end{table*}

We conduct an empirical study on model size to investigate the scalability of the {\modelname}. By progressively increasing the number of model parameters—specifically by expanding the hidden size and increasing the number of attention heads—we observe the changes in hierarchical recall (Hrecall@Level$l$@1000). Figure~\ref{fig:md_sz} presents the experimental results for \(l \in \{2, 3\}\) under different parameter scales. It can be observed that model performance improves steadily with scaling, demonstrating the scalability of the {\modelname} framework.
\subsection{Online A/B Test Results (RQ6)}

While improvements in offline evaluation metrics are observed, they do not necessarily translate into gains in real-world industrial settings. To validate the practical impact, we conducted online A/B tests with real users. We allocated 5\% of users as the control group, which was served recommendations from a generative recommendation model using only the recent 256 historical interactions. An additional 5\% of users were assigned to the experimental group, which received recommendations from the proposed model. The experiment ran for seven days on both the Kuaishou and Kuaishou Lite platforms. The results are presented in Table~\ref{tab:ab}.

{\modelname} achieves significant improvements across all key metrics in both products compared to the control group. On Kuaishou Lite, the \textbf{total app usage time increases by 0.17\%}, and the \textbf{total video watch time increases by 0.35\%}. These results indicate that our method effectively enhances user watching time, thereby improving user experience. This demonstrates that the proposed multi-stream decoder paradigm contributes to higher user engagement.

%% file: chapters/deploy.tex
\section{Training and Deployment Optimization}

\subsection{Training}
The model is trained using an online learning approach. We have implemented several optimizations to maximize training efficiency on modern high-performance GPUs. 1) \textbf{Mixed Precision} training with BF16/FP16 is employed, which increases training throughput on a single GPU. As a result, a larger batch size can be used (increased by \textasciitilde42\%), and training runtime improves by \textasciitilde37\%. This significantly enhances the efficiency of streaming training. 2) \textbf{Flash Attention} is adopted to accelerate attention computations. Given the extensive use of attention mechanisms in the model, this optimization further increases the batch size per GPU by  \textasciitilde10\%.

\subsection{Online Serving}
We have successfully deployed {\modelname} in an industrial setting, validating its availability in scenarios with strict inference latency constraints. 1) \textbf{Lightweight Indexer for Cross Attention} structure substantially reduces graph runtime. Since the model only needs to compute over a small subset of user sequence tokens, inference efficiency is greatly improved. By deploying the indexer, {\modelname} reduces latency by \textasciitilde70\% while maintaining comparable online inference performance. Table \ref{tab:ie} previously presents experimental comparisons between the indexer and brute-force aggregation. 2) \textbf{Mixed precision} inference is utilized to accelerate online serving, which reduces graph runtime by \textasciitilde21\%. 3) \textbf{TensorRT} is used to optimize the computation graph of {\modelname} in actual deployment, freeing \textasciitilde70\% of GPU memory. 4) \textbf{User Batching Scheme} is introduced to processe multiple user requests simultaneously during a single computation graph execution, further reducing latency by \textasciitilde12\%.

%% file: chapters/related.tex
\section{Related Work}
\subsection{Generative Recommendations}
Generative recommendation is steering the research direction of recommendation systems from traditional DNN architectures towards generative architectures. It is defined as a system where the model autoregressively predicts the next item a user will interact with, based on the user's historical sequence \cite{202512.0203}. Early pioneering works, such as DSI \cite{tay2022transformer} and TIGER \cite{rajput2023recommender}, employ the T5 architecture to encode user sequences and decode target Semantic ID sequences. LC-Rec \cite{zheng2024adapting} integrates collaborative signals into the training of generative models. COBRA enhances the incorporation of multimodal semantic information within generative recommendation. OneRec \cite{Zhou2025} proposes an end-to-end recommendation paradigm to replace the traditional cascaded framework. More recently, OneLoc \cite{wei2025oneloc} introduces a model for POI recommendation that conditions generation on geo-aware semantic information.
Beyond model architecture, discrete semantic tokenization is also a primary research focus in generative recommendation. TIGER quantizes items into discrete Semantic IDs based on RQ-VAE \cite{lee2022autoregressive}. QARM \cite{luo2025qarm} further employs Residual K-means to reduce codebook collision rates. LETTER \cite{wang2024learnable} incorporates hierarchical semantics, collaborative signals, and code assignment diversity into its tokenization process.

However, existing generative recommendation methods have not adequately addressed the modeling of lifelong sequences in real-world industrial scenarios. They fail to consider how to incorporate multi-scale sequential user features under the strict latency constraints typical of industrial environments.

\subsection{Lifelong Sequence Modeling}
Several ranking and retrieval models have explored methods for incorporating long user sequences. TWIN \cite{Chang2023} and its successor TWIN V2 \cite{Si2024} efficiently adapt information from sequences of up to 10,000 items within ranking models. LONGER \cite{chai2025longerscalinglongsequence} effectively introduces sequences of a similar length  into the retrieval stage. OneTrans \cite{zhang2025onetrans} introduces an end-to-end Transformer mechanism capable of modeling sequences with a length of 2048. Furthermore, MARM \cite{Lv2024} successfully models sequences of up to 6400 items based on a cached memory mechanism.

Within the domain of generative recommendation, user sequences typically remain under 10,000 items. For instance, Meta's HSTU \cite{zhai2024actionsspeaklouderwords} models sequences of 8,000 items within a Transformer-based architecture. Other models such as HLLM \cite{chen2024hllmenhancingsequentialrecommendations} handle approximately 1,000 items, while Dual-Flow \cite{guo2025action}, LUM \cite{yan2025unlockingscalinglawindustrial} manage 4,000 items, and GenRank \cite{huang2025towards} processes fewer than 500 items. It is evident that existing research, particularly in generative recommendation, has not sufficiently explored the efficient modeling of lifelong user sequences, especially those exceeding 100,000 interactions.

%% file: chapters/conclusion.tex
\section{Conclusion}

In this paper, we introduced {\modelname}, a multi-stream decoder framework that effectively extends the sequence modeling capacity of generative recommendations to the entire user lifecycle. By decoupling user interests into recent, mid-term, and lifelong streams with specialized encoding mechanisms, {\modelname} addresses the critical challenges of high computational costs and the "recency bias" inherent in long-sequence modeling. Experimental results and online A/B testing in a large-scale industiral scenario confirm that {\modelname} not only enhances recommendation performance by capturing multifaceted user interests but also meets the stringent latency requirements of industrial real-time systems. Our work provides a practical and efficient solution for integrating ultra-long user behavior modeling into the next generation of generative recommendation systems.

%% file: reference.bib
@String{Computer = "{IEEE} Computer" }

@article{zhang2025onetrans,
  title={OneTrans: Unified Feature Interaction and Sequence Modeling with One Transformer in Industrial Recommender},
  author={Zhang, Zhaoqi and Pei, Haolei and Guo, Jun and Wang, Tianyu and Feng, Yufei and Sun, Hui and Liu, Shaowei and Sun, Aixin},
  journal={arXiv preprint arXiv:2510.26104},
  year={2025}
}

@article{xu2025mmq,
  title={Mmq: Multimodal mixture-of-quantization tokenization for semantic id generation and user behavioral adaptation},
  author={Xu, Yi and Zhang, Moyu and Li, Chenxuan and Liao, Zhihao and Xing, Haibo and Deng, Hao and Hu, Jinxin and Zhang, Yu and Zeng, Xiaoyi and Zhang, Jing},
  journal={arXiv preprint arXiv:2508.15281},
  year={2025}
}

@Article{Zhou2025,
  author  = {Zhou, Guorui and Deng, Jiaxin and Zhang, Jinghao and Cai, Kuo and Ren, Lejian and Luo, Qiang and Wang, Qianqian and Hu, Qigen and Huang, Rui and Wang, Shiyao and others},
  journal = {arXiv preprint arXiv:2506.13695},
  title   = {OneRec Technical Report},
  year    = {2025},
}

@article{wei2025oneloc,
  title={Oneloc: Geo-aware generative recommender systems for local life service},
  author={Wei, Zhipeng and Cai, Kuo and She, Junda and Chen, Jie and Chen, Minghao and Zeng, Yang and Luo, Qiang and Zeng, Wencong and Tang, Ruiming and Gai, Kun and others},
  journal={arXiv preprint arXiv:2508.14646},
  year={2025}
}

@article{guo2025action,
  title={Action is All You Need: Dual-Flow Generative Ranking Network for Recommendation},
  author={Guo, Hao and Xue, Erpeng and Huang, Lei and Wang, Shichao and Wang, Xiaolei and Wang, Lei and Wang, Jinpeng and Chen, Sheng},
  journal={arXiv preprint arXiv:2505.16752},
  year={2025}
}

@inproceedings{luo2025qarm,
  title={Qarm: Quantitative alignment multi-modal recommendation at kuaishou},
  author={Luo, Xinchen and Cao, Jiangxia and Sun, Tianyu and Yu, Jinkai and Huang, Rui and Yuan, Wei and Lin, Hezheng and Zheng, Yichen and Wang, Shiyao and Hu, Qigen and others},
  booktitle={Proceedings of the 34th ACM International Conference on Information and Knowledge Management},
  pages={5915--5922},
  year={2025}
}

@inproceedings{lee2022autoregressive,
  title={Autoregressive image generation using residual quantization},
  author={Lee, Doyup and Kim, Chiheon and Kim, Saehoon and Cho, Minsu and Han, Wook-Shin},
  booktitle={Proceedings of the IEEE/CVF conference on computer vision and pattern recognition},
  pages={11523--11532},
  year={2022}
}

@Article{Zhou2025a,
  author  = {Zhou, Guorui and Hu, Hengrui and Cheng, Hongtao and Wang, Huanjie and Deng, Jiaxin and Zhang, Jinghao and Cai, Kuo and Ren, Lejian and Ren, Lu and Yu, Liao and others},
  journal = {arXiv preprint arXiv:2508.20900},
  title   = {Onerec-v2 technical report},
  year    = {2025},
}

@InProceedings{Han2025,
  author    = {Han, Ruidong and Yin, Bin and Chen, Shangyu and Jiang, He and Jiang, Fei and Li, Xiang and Ma, Chi and Huang, Mincong and Li, Xiaoguang and Jing, Chunzhen and others},
  booktitle = {Proceedings of the 34th ACM International Conference on Information and Knowledge Management},
  title     = {Mtgr: Industrial-scale generative recommendation framework in meituan},
  year      = {2025},
  pages     = {5731--5738},
}

@Article{Badrinath2025,
  author  = {Badrinath, Anirudhan and Agarwal, Prabhat and Bhasin, Laksh and Yang, Jaewon and Xu, Jiajing and Rosenberg, Charles},
  journal = {arXiv preprint arXiv:2504.10507},
  title   = {PinRec: Outcome-Conditioned, Multi-Token Generative Retrieval for Industry-Scale Recommendation Systems},
  year    = {2025},
}

@Article{Yi2025,
  author  = {Yi, Chao and Chen, Dian and Guo, Gaoyang and Tang, Jiakai and Wu, Jian and Yu, Jing and Zhang, Mao and Dai, Sunhao and Chen, Wen and Yang, Wenjun and others},
  journal = {arXiv preprint arXiv:2507.22879},
  title   = {Recgpt technical report},
  year    = {2025},
}

@InProceedings{Si2024,
  author    = {Si, Zihua and Guan, Lin and Sun, ZhongXiang and Zang, Xiaoxue and Lu, Jing and Hui, Yiqun and Cao, Xingchao and Yang, Zeyu and Zheng, Yichen and Leng, Dewei and others},
  booktitle = {Proceedings of the 33rd ACM International Conference on Information and Knowledge Management},
  title     = {Twin v2: Scaling ultra-long user behavior sequence modeling for enhanced ctr prediction at kuaishou},
  year      = {2024},
  pages     = {4890--4897},
}

@InProceedings{Chang2023,
  author    = {Chang, Jianxin and Zhang, Chenbin and Fu, Zhiyi and Zang, Xiaoxue and Guan, Lin and Lu, Jing and Hui, Yiqun and Leng, Dewei and Niu, Yanan and Song, Yang and others},
  booktitle = {Proceedings of the 29th ACM SIGKDD Conference on Knowledge Discovery and Data Mining},
  title     = {TWIN: TWo-stage interest network for lifelong user behavior modeling in CTR prediction at kuaishou},
  year      = {2023},
  pages     = {3785--3794},
}

@InProceedings{Guo2025,
  author    = {Guo, Chengcheng and She, Junda and Cai, Kuo and Wang, Shiyao and Hu, Qigen and Luo, Qiang and Zhou, Guorui and Gai, Kun},
  booktitle = {Proceedings of the 34th ACM International Conference on Information and Knowledge Management},
  title     = {MISS: Multi-Modal Tree Indexing and Searching with Lifelong Sequential Behavior for Retrieval Recommendation},
  year      = {2025},
  pages     = {5683--5690},
}

@Article{Lv2024,
  author  = {Lv, Xiao and Cao, Jiangxia and Guan, Shijie and Zhou, Xiaoyou and Qi, Zhiguang and Zang, Yaqiang and Li, Ming and Wang, Ben and Gai, Kun and Zhou, Guorui},
  journal = {arXiv preprint arXiv:2411.09425},
  title   = {MARM: Unlocking the Future of Recommendation Systems through Memory Augmentation and Scalable Complexity},
  year    = {2024},
}

@inproceedings{zheng2024full,
  title={Full stage learning to rank: A unified framework for multi-stage systems},
  author={Zheng, Kai and Zhao, Haijun and Huang, Rui and Zhang, Beichuan and Mou, Na and Niu, Yanan and Song, Yang and Wang, Hongning and Gai, Kun},
  booktitle={Proceedings of the ACM Web Conference 2024},
  pages={3621--3631},
  year={2024}
}

@article{liu2024kuaiformer,
  title={KuaiFormer: Transformer-Based Retrieval at Kuaishou},
  author={Liu, Chi and Cao, Jiangxia and Huang, Rui and Zheng, Kai and Luo, Qiang and Gai, Kun and Zhou, Guorui},
  journal={arXiv preprint arXiv:2411.10057},
  year={2024}
}

@inproceedings{zhu2018learning,
  title={Learning tree-based deep model for recommender systems},
  author={Zhu, Han and Li, Xiang and Zhang, Pengye and Li, Guozheng and He, Jie and Li, Han and Gai, Kun},
  booktitle={Proceedings of the 24th ACM SIGKDD international conference on knowledge discovery \& data mining},
  pages={1079--1088},
  year={2018}
}

@Article{Zhou2024,
  author    = {Zhou, Wen-Ji and Zheng, Yuhang and Feng, Yinfu and Ye, Yunan and Xiao, Rong and Chen, Long and Yang, Xiaosong and Xiao, Jun},
  journal   = {IEEE Transactions on Knowledge and Data Engineering},
  title     = {ENCODE: Breaking the Trade-Off Between Performance and Efficiency in Long-Term User Behavior Modeling},
  year      = {2024},
  publisher = {IEEE},
}

@article{202512.0203,
	doi = {10.20944/preprints202512.0203.v1},
	url = {https://doi.org/10.20944/preprints202512.0203.v1},
	year = 2025,
	month = {December},
	publisher = {Preprints},
	author = {Xiaopeng Li and Bo Chen and Junda She and Shiteng Cao and You Wang and Qinlin Jia and Haiying He and Zheli Zhou and Zhao Liu and Ji Liu and Zhiyang Zhang and Yu Zhou and Guoping Tang and Yiqing Yang and Chengcheng Guo and Si Dong and Kuo Cai and Pengyue Jia and Maolin Wang and Wanyu Wang and Shiyao Wang and Xinchen Luo and Qigen Hu and Qiang Luo and Xiao Lv and Chaoyi Ma and Ruiming Tang and Kun Gai and Guorui Zhou and Xiangyu Zhao},
	title = {A Survey of Generative Recommendation from a Tri-Decoupled Perspective: Tokenization, Architecture, and Optimization},
	journal = {Preprints}
}

@inproceedings{sun2025mpformer,
  title={MPFormer: Adaptive Framework for Industrial Multi-Task Personalized Sequential Retriever},
  author={Sun, Yijia and Huang, Shanshan and Che, Linxiao and Lu, Haitao and Luo, Qiang and Gai, Kun and Zhou, Guorui},
  booktitle={Proceedings of the 34th ACM International Conference on Information and Knowledge Management},
  pages={2832--2841},
  year={2025}
}

@misc{chai2025longerscalinglongsequence,
      title={LONGER: Scaling Up Long Sequence Modeling in Industrial Recommenders}, 
      author={Zheng Chai and Qin Ren and Xijun Xiao and Huizhi Yang and Bo Han and Sijun Zhang and Di Chen and Hui Lu and Wenlin Zhao and Lele Yu and Xionghang Xie and Shiru Ren and Xiang Sun and Yaocheng Tan and Peng Xu and Yuchao Zheng and Di Wu},
      year={2025},
      eprint={2505.04421},
      archivePrefix={arXiv},
      primaryClass={cs.IR},
      url={https://arxiv.org/abs/2505.04421}, 
}

@misc{guo2026promiseprocessrewardmodels,
      title={PROMISE: Process Reward Models Unlock Test-Time Scaling Laws in Generative Recommendations}, 
      author={Chengcheng Guo and Kuo Cai and Yu Zhou and Qiang Luo and Ruiming Tang and Han Li and Kun Gai and Guorui Zhou},
      year={2026},
      eprint={2601.04674},
      archivePrefix={arXiv},
      primaryClass={cs.IR},
      url={https://arxiv.org/abs/2601.04674}, 
}

@inproceedings{pi2019practice,
  title={Practice on long sequential user behavior modeling for click-through rate prediction},
  author={Pi, Qi and Bian, Weijie and Zhou, Guorui and Zhu, Xiaoqiang and Gai, Kun},
  booktitle={Proceedings of the 25th ACM SIGKDD international conference on knowledge discovery \& data mining},
  pages={2671--2679},
  year={2019}
}

@article{harper2015movielens,
  title={The movielens datasets: History and context},
  author={Harper, F Maxwell and Konstan, Joseph A},
  journal={Acm transactions on interactive intelligent systems (tiis)},
  volume={5},
  number={4},
  pages={1--19},
  year={2015},
  publisher={Acm New York, NY, USA}
}

@misc{yan2025unlockingscalinglawindustrial,
      title={Unlocking Scaling Law in Industrial Recommendation Systems with a Three-step Paradigm based Large User Model}, 
      author={Bencheng Yan and Shilei Liu and Zhiyuan Zeng and Zihao Wang and Yizhen Zhang and Yujin Yuan and Langming Liu and Jiaqi Liu and Di Wang and Wenbo Su and Wang Pengjie and Jian Xu and Bo Zheng},
      year={2025},
      eprint={2502.08309},
      archivePrefix={arXiv},
      primaryClass={cs.IR},
      url={https://arxiv.org/abs/2502.08309}, 
}

@inproceedings{mcauley2015image,
  title={Image-based recommendations on styles and substitutes},
  author={McAuley, Julian and Targett, Christopher and Shi, Qinfeng and Van Den Hengel, Anton},
  booktitle={Proceedings of the 38th international ACM SIGIR conference on research and development in information retrieval},
  pages={43--52},
  year={2015}
}

@article{huang2025towards,
  title={Towards Large-scale Generative Ranking},
  author={Huang, Yanhua and Chen, Yuqi and Cao, Xiong and Yang, Rui and Qi, Mingliang and Zhu, Yinghao and Han, Qingchang and Liu, Yaowei and Liu, Zhaoyu and Yao, Xuefeng and others},
  journal={arXiv preprint arXiv:2505.04180},
  year={2025}
}

@misc{chen2024hllmenhancingsequentialrecommendations,
      title={HLLM: Enhancing Sequential Recommendations via Hierarchical Large Language Models for Item and User Modeling}, 
      author={Junyi Chen and Lu Chi and Bingyue Peng and Zehuan Yuan},
      year={2024},
      eprint={2409.12740},
      archivePrefix={arXiv},
      primaryClass={cs.IR},
      url={https://arxiv.org/abs/2409.12740}, 
}

@inproceedings{wang2024learnable,
  title={Learnable item tokenization for generative recommendation},
  author={Wang, Wenjie and Bao, Honghui and Lin, Xinyu and Zhang, Jizhi and Li, Yongqi and Feng, Fuli and Ng, See-Kiong and Chua, Tat-Seng},
  booktitle={Proceedings of the 33rd ACM International Conference on Information and Knowledge Management},
  pages={2400--2409},
  year={2024}
}

@inproceedings{zheng2024adapting,
  title={Adapting large language models by integrating collaborative semantics for recommendation},
  author={Zheng, Bowen and Hou, Yupeng and Lu, Hongyu and Chen, Yu and Zhao, Wayne Xin and Chen, Ming and Wen, Ji-Rong},
  booktitle={2024 IEEE 40th International Conference on Data Engineering (ICDE)},
  pages={1435--1448},
  year={2024},
  organization={IEEE}
}

@article{tay2022transformer,
  title={Transformer memory as a differentiable search index},
  author={Tay, Yi and Tran, Vinh and Dehghani, Mostafa and Ni, Jianmo and Bahri, Dara and Mehta, Harsh and Qin, Zhen and Hui, Kai and Zhao, Zhe and Gupta, Jai and others},
  journal={Advances in Neural Information Processing Systems},
  volume={35},
  pages={21831--21843},
  year={2022}
}

@inproceedings{li2019multi,
  title={Multi-interest network with dynamic routing for recommendation at Tmall},
  author={Li, Chao and Liu, Zhiyuan and Wu, Mengmeng and Xu, Yuchi and Zhao, Huan and Huang, Pipei and Kang, Guoliang and Chen, Qiwei and Li, Wei and Lee, Dik Lun},
  booktitle={Proceedings of the 28th ACM international conference on information and knowledge management},
  pages={2615--2623},
  year={2019}
}

@misc{zhai2024actionsspeaklouderwords,
      title={Actions Speak Louder than Words: Trillion-Parameter Sequential Transducers for Generative Recommendations}, 
      author={Jiaqi Zhai and Lucy Liao and Xing Liu and Yueming Wang and Rui Li and Xuan Cao and Leon Gao and Zhaojie Gong and Fangda Gu and Michael He and Yinghai Lu and Yu Shi},
      year={2024},
      eprint={2402.17152},
      archivePrefix={arXiv},
      primaryClass={cs.LG},
      url={https://arxiv.org/abs/2402.17152}, 
}

@article{liu2024crm,
  title={CRM: Retrieval Model with Controllable Condition},
  author={Liu, Chi and Cao, Jiangxia and Huang, Rui and Cai, Kuo and Ding, Weifeng and Luo, Qiang and Gai, Kun and Zhou, Guorui},
  journal={arXiv preprint arXiv:2412.13844},
  year={2024}
}

@article{sun2025grank,
  title={GRank: Towards Target-Aware and Streamlined Industrial Retrieval with a Generate-Rank Framework},
  author={Sun, Yijia and Huang, Shanshan and Guan, Zhiyuan and Luo, Qiang and Tang, Ruiming and Gai, Kun and Zhou, Guorui},
  journal={arXiv preprint arXiv:2510.15299},
  year={2025}
}

@article{rajput2023recommender,
  title={Recommender systems with generative retrieval},
  author={Rajput, Shashank and Mehta, Nikhil and Singh, Anima and Hulikal Keshavan, Raghunandan and Vu, Trung and Heldt, Lukasz and Hong, Lichan and Tay, Yi and Tran, Vinh and Samost, Jonah and others},
  journal={Advances in Neural Information Processing Systems},
  volume={36},
  pages={10299--10315},
  year={2023}
}

@article{chen2025massive,
  title={Massive Memorization with Hundreds of Trillions of Parameters for Sequential Transducer Generative Recommenders},
  author={Chen, Zhimin and Zhao, Chenyu and Mo, Ka Chun and Jiang, Yunjiang and Lee, Jane H and Chen, Shouwei and Mahajan, Khushhall Chandra and Jiang, Ning and Ren, Kai and Li, Jinhui and others},
  journal={arXiv preprint arXiv:2510.22049},
  year={2025}
}
